\documentclass[useAMS,usenatbib]{mn2e}
\usepackage{epsfig}
\usepackage{graphicx}
\usepackage{psfrag}
\newcommand{\mnras}{Monthly Notices of the Royal Astronomical Society}
\newcommand{\aj}{Astronomical Journal}
\newcommand{\apj}{Astrophysical Journal}
\newcommand{\apjl}{Astrophysical Journal Letters}
\newcommand{\aap}{Astronomy and Astrophysics}
\newcommand{\araa}{Annual Review of Astronomy and Astrophysics}
\newcommand{\aaps}{A\&AS}

\title[]{
Limitations of model fitting methods for lensing shear estimation
}
\author[L.M. Voigt and S. L. Bridle]{L. M. Voigt$^{1}$\thanks{E-mail:
lvoigt@star.ucl.ac.uk (LMV); sarah.bridle@ucl.ac.uk (SLB)} and S. L.
Bridle$^{1}$\footnotemark[1]\\
$^{1}$Department of Physics and Astronomy, University College London, UK}

\begin{document}

\date{Accepted . Received ; in original form }

\pagerange{\pageref{firstpage}--\pageref{lastpage}} \pubyear{2008}

\maketitle

\label{firstpage}

\begin{abstract}
Gravitational lensing shear has the potential to be the most powerful tool for constraining the nature of dark energy.
However, accurate measurement of galaxy shear is crucial and has been shown to be non-trivial by the Shear TEsting Programme.
Here we demonstrate a fundamental limit to the accuracy achievable by model-fitting techniques, if oversimplistic models are used.
We show that even if galaxies have elliptical isophotes, model-fitting methods which assume elliptical isophotes
can have significant biases if they use the wrong profile. We use noise-free simulations to show that on allowing
sufficient flexibility in the profile the biases can be made negligible.
This is no longer the case if elliptical isophote models are used to fit galaxies made up of a bulge plus a disk,
if these two components have different ellipticities. The limiting accuracy is dependent on the galaxy shape but we find the most significant biases for simple spiral-like galaxies. The implications for a given cosmic shear survey will depend on the actual distribution of galaxy morphologies in the universe, taking into account the survey selection function and the point spread function.
However our results suggest that the impact on cosmic shear results from current and near future surveys may be negligible.
Meanwhile, these results should encourage the development of existing approaches which are less sensitive to morphology, as well as methods which use priors
on galaxy shapes learnt from deep surveys.
\end{abstract}
\begin{keywords}
galaxy shapes
\end{keywords}

\section{Introduction}
Dark energy dominates the mass-energy of the universe and the goal to discover the nature of dark energy, or even whether it truly exists, is of paramount importance in cosmology. Cosmic shear provides one of the most promising methods for constraining the nature of dark energy~\citep{detf,esoesa}. Cosmic shear is the mild distortion of distant galaxy images due to the bending of light by intervening matter.
Typically galaxy images are stretched by only a few per cent, for example an intrinsically circular galaxy image would become an ellipse with major to minor axis ratio of about 1.06.
The clumpier the intervening dark matter, the greater the distortions. Dark energy affects the rate of gravitational collapse, therefore it can be investigated by measuring cosmic shear at different times in the history of the universe.

A number of observational surveys are planned to capitalise on this, including ground-based projects KIlo-Degree Survey (KIDS), Pan-STARRS~\footnote{http://pan-starrs.ifa.hawaii.edu}, the Dark Energy Survey (DES)~\footnote{http://www.darkenergysurvey.org} and the Large Synoptic Survey Telescope (LSST)~\footnote{http://www.lsst.org}, and space missions the International Dark Energy Cosmology Survey (IDECS) or Euclid and/or the Joint Dark Energy Mission (JDEM). If we are to fully utilise the potential of these future cosmology surveys then the potential systematics associated with measuring cosmic lensing must be understood and controlled. The main areas for work are (i) measurement and calibration of galaxy redshifts (ii) measurement and subtraction of galaxy intrinsic alignments and (iii) accurate shear measurement from images. In this paper we focus on the last of these.

Shear measurement is difficult because (i) images are convolved with a kernel due to the atmosphere, telescope optics and measurement devices, (ii) they are then pixelised and (iii) they are noisy mainly due to the finite number of photons collected. The convolution kernel (usually referred to as the point spread function, hereafter PSF) is typically a similar size to the unconvolved galaxy image and is generally not circular. It must be accurately measured either from a detailed model of the telescope or, more usually, from stars in the image, which can be treated as point objects before the convolution. Many works, including this paper, focus on the case where the PSF is perfectly known. However the shear measurement problem is still very difficult due to the high noise levels in the images and the very small signals that need to be measured. The signal-to-noise ratio on shear measurement from any single galaxy image is typically about 0.1, and the signal from many millions of galaxies must be combined to make useful measurements of cosmology.

The Shear TEsting Programme~\citep{heymansea04,STEP2} is a collaborative effort to quantify the biases associated with current shear measurement methods. Crucially, the programme has validated the implementation of the~\citet{kaisersb95} (KSB) method by several groups to obtain shears from real data. In brief, the method measures the quadrupole moments of the image which are combined to estimate the ellipticity of the galaxy. The presence of noise in the images requires the addition of a weighting factor. It is now widely believed however that KSB methods will not be sufficiently accurate to obtain shears from future surveys observing billions of galaxies.

Several groups are working on model-fitting methods to obtain shear, using either Gaussian weighted Hermite polynomials (`shapelets') to model the galaxy \citep{bernsteinj02,nakajimab07} or elliptical profiles \citep{kuijken99,bridlekbg02,Irwin:2006tn,kuijken06,millerkhhv07,kitchingmhvh08}. Alternatively, statistics from shapelets can also be considered as shear estimators that generalise and improve on weighted quadrupole moments~\citep{refregier03,refregierb03,masseyr05}. The GRavitational lEnsing Accuracy Testing 2008 (GREAT08) Challenge~\citep{great08} has recently been run to draw expertise from researchers in statistical inference, inverse problems and computational learning.

There is a large variety of galaxy morphologies, whereas the amount of information in any single typical galaxy image is extremely small. Model-fitting methods must therefore make some assumptions. \citet{lewisclt09} has shown that both the PSF and the galaxy shape must be accurately modelled to remove biases; in particular the paper proves that this is a direct result of the symmetries broken by the PSF. In this paper we concentrate solely on the galaxy model, quantifying the bias on the shear for models using elliptical profiles, and assume the PSF is known precisely. In addition we assume infinite signal-to-noise. We first consider the case where the simulated galaxy also has elliptical isophotes, adopting the widely-used de Vaucouleurs and exponential profiles. We also consider more realistic simulated galaxies with non-elliptical isophotes, in particular two-component systems representing early (elliptical) and late-type (spiral or disk-dominated) galaxies in which each component has a different profile shape and ellipticity.

The paper is organised as follows. In Section~\ref{sect:shear_est} we summarise the equations governing gravitational shear and describe the method used to quantify the accuracy of the shear measurement method. In addition we discuss the requirements on the accuracy for future dark energy surveys. In Section~\ref{sect:simulations} we describe the simulations used to test the method and in Section~\ref{sect:model} we describe the shape measurement method. We then present results for different galaxy shapes in Sections~\ref{sect:isophotes} and~\ref{sect:bd}. Finally we discuss the implications of these results on the development of future methods in Section~\ref{sect:discussion}.

\section[]{Shear estimation}
\label{sect:shear_est}

\subsection[]{Gravitational shear}
\label{sect:grav_shear}
Light from a source passing a thin lens at position $\btheta$ in the lens plane suffers a deflection through an angle $\balpha$ given by
\begin{eqnarray}
\balpha = \nabla \Psi(\btheta),
\label{eqn:one}
\end{eqnarray}
where $\Psi$ is the projected gravitational potential of the lens. If $\bbeta$ is the true position of the source then the observed position $\btheta$ is related to $\balpha$ through the lens equation
\begin{eqnarray}
\balpha(\btheta) = \btheta - \bbeta.
\label{eqn:two}
\end{eqnarray}
The gravitational potential of the lens at $\btheta$ is related to its surface mass density $\Sigma(\btheta)$ via the Poisson equation
\begin{eqnarray}
\nabla^{2} \Psi(\btheta) = \frac{\Sigma(\btheta)}{\Sigma_{\rm crit}} = 2 \kappa(\btheta),
\label{eqn:three}
\end{eqnarray}
where $\kappa(\btheta)$ is the convergence and the critical surface density is
\begin{eqnarray}
\Sigma_{\rm crit} = \frac{c^2}{4 \pi G} \frac{D_{ls}}{D_{l}D_{s}},
\end{eqnarray}
where $D_{s}, D_{l}$ and $D_{ls}$ are the angular-diameter distances between the observer and the source, the observer and the lens and between the lens and the source, respectively.

Differentiating Eqns.~\ref{eqn:one} and ~\ref{eqn:two} with respect to $\btheta$ we obtain the Jacobian, or magnification matrix, relating the apparent position $\btheta$ to the unlensed position $\bbeta$ in terms of the gradients of the gravitational potential
\begin{equation}
\emph{\textbf{M}}=\frac{\partial\bbeta}{\partial\btheta}=\left(
                    \begin{array}{cc}
                      1-\psi_{11} & -\psi_{12} \\
                      -\psi_{21} & 1-\psi_{22} \\
                    \end{array}
                  \right)
\end{equation}
where \(\psi_{ij}=\partial^{2}\psi/\partial \theta_{i} \partial \theta_{j}\).

Defining the complex gravitational shear as
\begin{equation}
\gamma=\gamma_{1}+i\gamma_{2}, %S20090526 [bgamma implies a vector, but gamma is not a vector, just a complex number. Could use vectors *instead* of complex notation e.g. bgamma = {gamma_1, gamma_2} but we haven't done this. Ditto for the ellipticities. Even if you personally prefer bold for complex numbers, MNRAS will change it back to ordinary, since it reserves bold for vectors. I guess could keep it different on astroph but I prefer if we stick to convention.]
\end{equation}
\label{eqn:gamma}
with
\begin{equation}
\gamma_{1}=\frac{1}{2}
\left(\psi_{11}-\psi_{22}\right),
\gamma_{2}=\psi_{12}=\psi_{21},
\end{equation}
the magnification matrix becomes
\begin{equation}
\emph{\textbf{M}}=\left(
                    \begin{array}{cc}
                      1-\kappa-\gamma_{1} & -\gamma_{2} \\
                      -\gamma_{2} & 1-\kappa+\gamma_{1} \\
                    \end{array}
                  \right).
\end{equation}
Under this transformation an object with intrinsic complex ellipticity given by
\begin{equation}
{e}^{\mathrm{s}}
=\frac{a-b}{a+b}
e^{2i\phi},
\end{equation}
\label{eqn:defe}
where $a$ and $b$ are the major and minor axes and $\phi$ is the orientation
of the major axis from the $x$-axis, is sheared to an object with observed complex ellipticity,
$e^{\mathrm{o}}$,
given by
\begin{equation}
e^{\mathrm{o}}=\frac{e^{\mathrm{s}}+g}{1+g^{*}e^{\mathrm{s}}}
\label{eqn:eobs}
\end{equation}
\citep{seitzs97}, where
$g=\gamma/(1-\kappa)$
is the reduced shear and we have assumed $|g|<1$ (which applies throughout this paper).

\subsection[]{Quantifying the bias on the shear estimator}
\label{sect:bias_shear}

Shape noise is the statistical noise arising from the random distribution of galaxy shapes. We quantify the bias on the shear measured for different galaxy shapes in the absence of shape noise (i.e. in the limit of an infinite number of galaxy orientations). To achieve this we follow~\citet{nakajimab07} by performing a `ring-test', whereby the same galaxy is rotated around a ring prior to shearing.
The mean ellipticity over the ring provides a shear estimate which, as explained below, is free from shape noise to first order.
Our shear estimator, $\hat{\gamma}$,
is the measured galaxy ellipticity,
$e^{\mathrm{m}}$. For a perfect shear measurement method the measured ellipticity is equal to the true observed ellipticity, given in
Eq.~\ref{eqn:eobs}.
Even in this case, averaging over galaxy orientations $i$ gives the following expression for the mean true observed ellipticity
\begin{equation}
\langle
e^{\mathrm{o}}_i %S20090526
%S20090526 \emph{\textbf{e}}^{\mathrm{o}}_i %SLB1404 [added subscript "i"] %LMV2505 [changed e to bold]
\rangle =
%S20090526 \bgamma^{\mathrm{t}}+
\gamma^{\mathrm{t}}+ %S20090526
\langle
e^{\mathrm{s}}_{i} %S20090526
%S20090526 \emph{\textbf{e}}^{\mathrm{s}}_{i} %LMV2505
\rangle +
\langle
(e^{\mathrm{s}}_{i}+\gamma^{\mathrm{t}})(-\gamma^{\mathrm{t}*}e^{\mathrm{s}}_{i}+O(\gamma^{\mathrm{t}*}e^{\mathrm{s}}_{i}))
%S20090526 (\emph{\textbf{e}}^{\mathrm{s}}_{i}+\bgamma^{\mathrm{t}})(-\bgamma^{\mathrm{t}*}\emph{\textbf{e}}^{\mathrm{s}}_{i}+O(\bgamma^{\mathrm{t}*}\emph{\textbf{e}}^{\mathrm{s}}_{i}))
\rangle
\label{eqn:eobs_mean}
\end{equation}
where
$\gamma^{\mathrm{t}}$ %S20090526
is the true input shear. The term
$\langle e^{\mathrm{s}}_{i} \rangle$ %SLB1404
is zero for a pair of identical galaxies rotated by 90 degrees from each other. Measuring biases for galaxy pairs was suggested by~\citet{nakajimab07} and adopted in the STEP2 simulations~\citep{Massey:2006ha} as a useful method for reducing the intrinsic shape noise.
We find that using three linearly spaced pairs of galaxies in the ring-test is enough to reduce the total contribution to the shape noise (i.e. including higher order terms in the sum in Eqn~\ref{eqn:eobs_mean}) to a negligible level.
To test the effects of PSF convolution and pixellisation on the accuracy of our (non-perfect) shear measurement method (i.e. in which
$e^{\mathrm{m}} \neq e^{\mathrm{o}}$)
we use 18 linearly spaces angles between 0 and 170 degrees. We find that the biases on the shear measured do not change if we double the number of angles used.

We quantify the bias on the shear estimator in terms of multiplicative and additive errors, $m_i$  and $c_i$ respectively, following~\citet{STEP1}, such that
\begin{equation}
\hat{\gamma}_i=(1+m_i) \gamma^{\rm t}_i+c_i
\end{equation}
where we assume there is no cross contamination of e.g. $\hat{\gamma_1}$ depending on the value of $\gamma^{\rm t}_2$ or vica versa.
We measure $m_1$ ($m_2$) by shearing along (at 45$^{\circ}$ to) the $x$-axis with a magnitude of 0.03, i.e. we measure $m_1$ by shearing using $\gamma^{\rm t}_1=0.03$, $\gamma^{\rm t}_2=0$ and $m_2$ by shearing using $\gamma^{\rm t}_1=0$, $\gamma^{\rm t}_2=0.03$.

\subsection[]{Bias requirements for future surveys}
\label{sect:requirements}

\citet{amara08} derived requirements on $m_i$ and $c_i$ for general
current and future surveys covering $A$ deg$^{2}$ of sky, with $n_{\rm gal}$ galaxies per arcmin$^{2}$ and with a median redshift $z_{\rm m}$
(their Eqns. 21 and 22). They consider general functional forms for the redshift evolution of these parameters and require the systematic biases from shear calibration to be less than the random uncertainties, for a two-parameter dark energy equation of state.

We consider 3 sets of survey parameters ($A$,~$n_{\rm gal}$,~$z_{\rm m}$): (170,~12,~0.8), (5000,~12,~0.8) and ($2\times10^{4}$,~35,~0.9).
These parameter sets are chosen to represent the Canada-France-Hawaii Telescope Legacy Survey (CFHTLS), the Dark Energy Survey (DES) and Euclid. We assume the limit on the additive error $c_i$ is equal to the limit on $\sigma_{\rm sys}$ in their
Eq.~21 therefore this gives the limits given in Table~\ref{tab:req}
for each of the three fiducial surveys.

\begin{table}
\center
\caption{Shear measurement requirements on current and future surveys, rounded to one decimal place. For specific survey parameters area $A$, number of galaxies per square arcminute $n_{\rm gal}$ and median redshift $z_{\rm m}$ we show the requirements on the shear multiplicative bias parameters $m_i$ and the shear additive bias parameter $c_i$. These are combined using the assumptions detailed around Eq.~\ref{eqn:q} to estimate a required GREAT08 $Q$ value. Survey parameters for upcoming, mid-term and far-future surveys are inspired by the CFHTLS Legacy Survey, the Dark Energy Survey (DES) and Euclid.}
\label{tab:req}
\begin{tabular}{|c|ccc|ccc|}
\hline
Survey &  $A$ & $n_{\rm gal}$ &  $z_{\rm m}$ & $m_i$ & $c_i$ & $Q$ \\
\hline
Upcoming & 170 & 12 & 0.8 & 0.02 & 0.001 & 43\\
Mid-term & 5000 & 12 & 0.8 & 0.004 & 0.0006 & 260\\
Far-future & $2 \times 10^{4} $ &  35 & 0.9 & 0.001 & 0.0003 & 990\\
\hline
\end{tabular}
\label{tab:requirements}
\end{table}

The GRavitational lEnsing Accuracy Testing 2008 (GREAT08) Challenge~\citep{great08} has set a target accuracy level, described by the quality factor, $Q$. The quality factor can be related to the $m$ and $c$ values via the equation
\begin{equation}
Q=\frac{10^{-4}}{ \langle m_{i}^{2} \sigma_{\gamma}^2 + c_{i}^{2} \rangle_i },
\label{eqn:q}
\end{equation}
where $i$ refers to the two shear components and we have written $\sigma_{\gamma}$ as the rms shear used in the simulation (technically this is the reduced shear rather than the shear) and we have assumed that $m_i$ and $c_i$ are the same for all data. We further assumed that the mean true shear in the simulation is zero. Typically $\sigma_{\gamma}\sim 0.03$ for cosmic shear. The GREAT08 Challenge has set a target accuracy level of $Q=1000$. Therefore, if $m_i=0$ then this corresponds to the Euclid requirement on $c_i$.

\section[]{Simulations}
\label{sect:simulations}

We next describe the simulations we have used to investigate biases in shear measurement. In Section~\ref{sect:isophotes} we investigate shear measurement from simulated de Vaucouleurs and exponential profiles, and in Section~\ref{sect:bd} two-component galaxies in which each component has a different S\'{e}rsic index and ellipticity.
Therefore here we discuss the two different galaxy profiles considered, the method used for convolution and the two-component models.

\subsection[]{Galaxy profiles}
\label{sect:obs_profiles}
Galaxies are broadly classified in the literature as ellipticals, pure spheroids or spheroid (bulge) plus disk systems. The de Vaucouleurs profile has long been used to model the light from elliptical galaxies~\citep{devau1948} and the exponential profile provides a good description of disk galaxies both in the local universe~\citep{freeman1970,kormendy1977,jong1996,macarthur03} and at high redshift~\citep{elmegreen2005}. Historically, pure spheroids and bulge components have also been modelled using a de Vaucouleurs profile, though recent studies have revealed a range of profile shapes \citep{GrahamWorley}.

Both the de Vaucouleurs and exponential profiles belong to a family of functions known as the S\'{e}rsic profiles \citep{sersic1968}. The S\'{e}rsic intensity at position $\emph{\textbf{x}}$ is given by
\begin{eqnarray}
I(\emph{\textbf{x}})=A e^{-k [(\emph{\textbf{x}}-\emph{\textbf{x}}_{0})^T \emph{\textbf{C}} (\emph{\textbf{x}}-\emph{\textbf{x}}_{0})]^{\frac{1}{2n}}}
\end{eqnarray}
where $\emph{\textbf{x}}_{0}$ is the centre, $A$ is the peak intensity, $n$ is the S\'{e}rsic index and
%LMV2605 $C$
$\emph{\textbf{C}}$ (proportional to the inverse covariance matrix
if $n=0.5$)
has elements
\begin{eqnarray}
C_{11}=\left(\frac{\rm cos^{2} \phi}{a^2}+\frac{\rm sin^{2} \phi}{b^2}\right)
\end{eqnarray}
\begin{eqnarray}
C_{12}=\frac{1}{2}\left(\frac{1}{a^2}-\frac{1}{b^2}\right) \rm{sin} (2\phi)
\end{eqnarray}
\begin{eqnarray}
C_{22}=\left(\frac{\rm sin^{2} \phi}{a^2}+\frac{\rm cos^{2} \phi}{b^2}\right)
\end{eqnarray}
where $\phi$ is the angle (measured anti-clockwise) between the $x$-axis and the major axis of the ellipse and the minor to major axis ratio is $b/a$.
The S\'{e}rsic index defines the profile `type', with $n=0.5, 1$ and $4$ for Gaussian, exponential and de Vaucouleurs profiles respectively. If $k$ is defined as
$k=1.9992n-0.3271$ then for a circular profile $r_{\rm e}=a=b$, referred to as the `effective radius' or `half-light radius', is the radius enclosing half the total flux.
(Note that for a Gaussian profile $a^{2}$ and $b^{2}$ are the 2D variances if $k=0.5$;
for the exponential profile $h=a=b$ is known as the `scale length' when $k=1$.)
The full width at half-maximum intensity (FWHM) is related to the effective radius (for a circular profile) via
\begin{eqnarray}
\mathrm{FWHM}=2r_{\mathrm{e}}\left(\frac{\ln 2}{
k
}\right)^{n}.
\end{eqnarray}
Thus for similar effective radii the FWHM of galaxies modelled by S\'{e}rsic indices ranging between 0.5 and 4 vary by nearly 4 orders of magnitude.
The total flux (integrated to infinity) emitted by a
galaxy described by a S\'{e}rsic profile with index $n$ is given by
\begin{equation}
F=2\pi n k^{-2n}r_{\rm e}^{2} \Gamma(2n) I_{0}
\label{eqn:flux}
\end{equation}
where $\Gamma$ is the gamma function.

\subsection[]{Shear and Convolution}
\label{sect:sim_gal}

\begin{figure}
\center
\epsfig{file=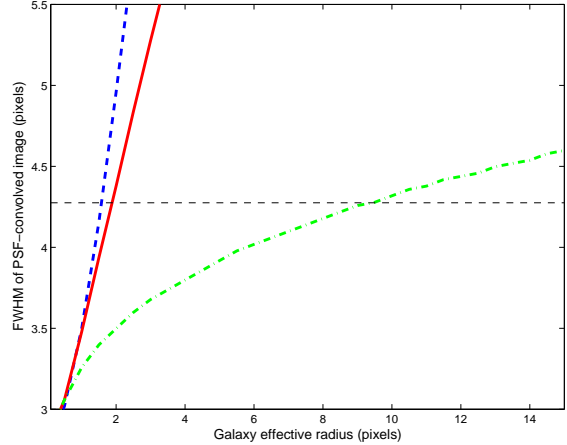,height=6cm,angle=0}
\caption{
Relationship between the FWHM of the PSF-convolved image and the effective radius of the galaxy. Curves are shown for Gaussian (blue dashed), exponential (red solid) and de Vaucouleurs (green dash-dot) profiles. The horizontal curve (black dashed) shows the FWHM of the PSF-convolved galaxy image used in this paper.
}
\label{fig:fwhm}
\end{figure}

We model the PSF  as a single Gaussian aligned along the $x$-axis with ellipticity
$e_{\rm p}=0.05$ and FWHM of 2.85 pixels. We define the FWHM of an elliptical object such that the area of the ellipse is equal to the area of a circle with the same FWHM.
The default value used for the galaxy ellipticity is $e=0.2$. %LMV2705
The galaxy size is chosen such that the FWHM  of the PSF-convolved image is
%S20090526 [I think no need to say approximately, given that you specify the exact method in the footnote.] approximately %LMV2704
1.5 times that of the PSF
\footnote{Specifically, we first compute the FWHM of the galaxy for the case where both the PSF and the galaxy are circular and the FWHM of the PSF-convolved image is 1.5 times that of the PSF. We then adjust the FWHM of the galaxy to keep the area of the ellipse constant as the ellipticity is increased.}. %LMV1905
Galaxies smaller than this are generally cut from catalogues used in weak lensing analyses.

We use Eqn.~\ref{eqn:eobs} to calculate the ellipticity and orientation of the sheared galaxy at each point in the ring. The major axis of the ellipse is held constant at the pre-lensed value and the minor axis adjusted to obtain the correct, post-shear ellipticity. For bulge plus disk galaxies we shear each component separately.

Fig~\ref{fig:fwhm} shows the relationship between the galaxy effective radius and the FWHM of the PSF-convolved image for Gaussian, de Vaucouleurs and exponential profiles.
The horizontal dashed line shows the value used in this study. Fig.~\ref{fig:oned} shows cross-sections through the galaxy and PSF-convolved galaxy profiles for the chosen galaxy parameters, compared with a Gaussian galaxy image. We see that the de Vaucouleurs has an extremely sharp galaxy profile before the PSF convolution, and larger wings after convolution.

By default the galaxy is convolved numerically with the PSF on a large, fine grid $(25\times45)^2$ pixels in size. The PSF FWHM is sampled by $(2.85\times45)$ pixels. Following the convolution the grid is binned up by a factor of 45 to obtain a square image 25 pixels across in which the FWHM of the PSF is 2.85 pixels. Finally, we cut the grid down to obtain a postage stamp 15 pixels across. We try increasing the resolution used for the convolution such that the PSF FWHM is sampled by $(2.85\times55)$ pixels. The grid is $(25\times55)^2$ pixels in size and, following the convolution, is binned up by a factor of 55. We also try increasing the size of the grid used for the convolution to $(31\times45)^2$ pixels, keeping the PSF FWHM at the default value and binning up by a factor of 45. In both cases it is the central $15^2$ pixels which are analysed. We find that the results do not change when we increase either the resolution or the grid size used for the convolution.

The true galaxy centroid is at the centre of the postage stamp. We find the results are largely insensitive to changes in the centroid position within the central pixel.

\subsection[]{Two-component models}
\label{sec:two_compt_models}

As discussed above, the de Vaucouleurs and exponential profiles are widely used to describe the light distribution in elliptical and disk galaxies.
However, real galaxies do not have constant ellipticity isophotes.
Therefore in this paper we also explore galaxies with both a bulge and a disk component and, crucially, with non-constant ellipticity isophotes, since we allow the bulge and disk to have different ellipticities.

We consider two different two-component systems: one which closely models realistic spiral (disk-dominated) galaxies, and one which represents ellipticals with a small disk (exponential) component. For the spiral galaxies the bulge is modelled as a S\'{e}rsic profile with index 1.5. While for many years it was believed that bulges were universally described by the $r^{1/4}$ model \citep{devau1948,devau1958,devau1978}, it is now generally accepted that most bulges have S\'{e}rsic indices $n<4$ \citep{Graham01,macarthur03,balcells03,Laurikainen06,GrahamWorley} and typically between $\sim 1-2$ for a range of Hubble types \citep[][see their figure 3]{GrahamWorley}. Studies also suggest that the bulge-to-disk size ratio is reasonably independent of Hubble type, with \citet{GrahamWorley} finding a median value for $r_{e}/h$ equal to 0.22. We adopt a similar size ratio, with
$r_{\rm d}/r_{\rm b}$
equal to 7.5,
where $r_{\rm d}$ and $r_{\rm b}$ are the disk and bulge effective radii, respectively.
Our second model is chosen to represent ellipticals, which are well-described by de Vaucouleurs profiles. We add a small exponential component such that
$r_{\rm d}/r_{\rm b}$=0.5.
For both models the bulge and disk ellipticities are set equal to
$e_{\rm b}=0.05$
and
$e_{\rm d}=0.2$
respectively. The bias on the shear is measured for a range of bulge-to-total ($B/T$) flux ratios between 0 and 1.
At each $B/T$ value the ratio
$r_{\rm d}/r_{\rm b}$
is held constant and
the bulge and disk effective radii
computed for circular PSF and galaxy profiles
such that the
FWHM
of the PSF-convolved image is 1.5 times the
FWHM
of the PSF.
The total flux in each galaxy component is calculated by integrating the flux from $r=0$ to infinity, as given in Eqn.~\ref{eqn:flux}. The bulge and disk effective radii are then adjusted to keep the area of each component constant as the ellipticity is increased from zero.

\begin{figure}
\center
\epsfig{file=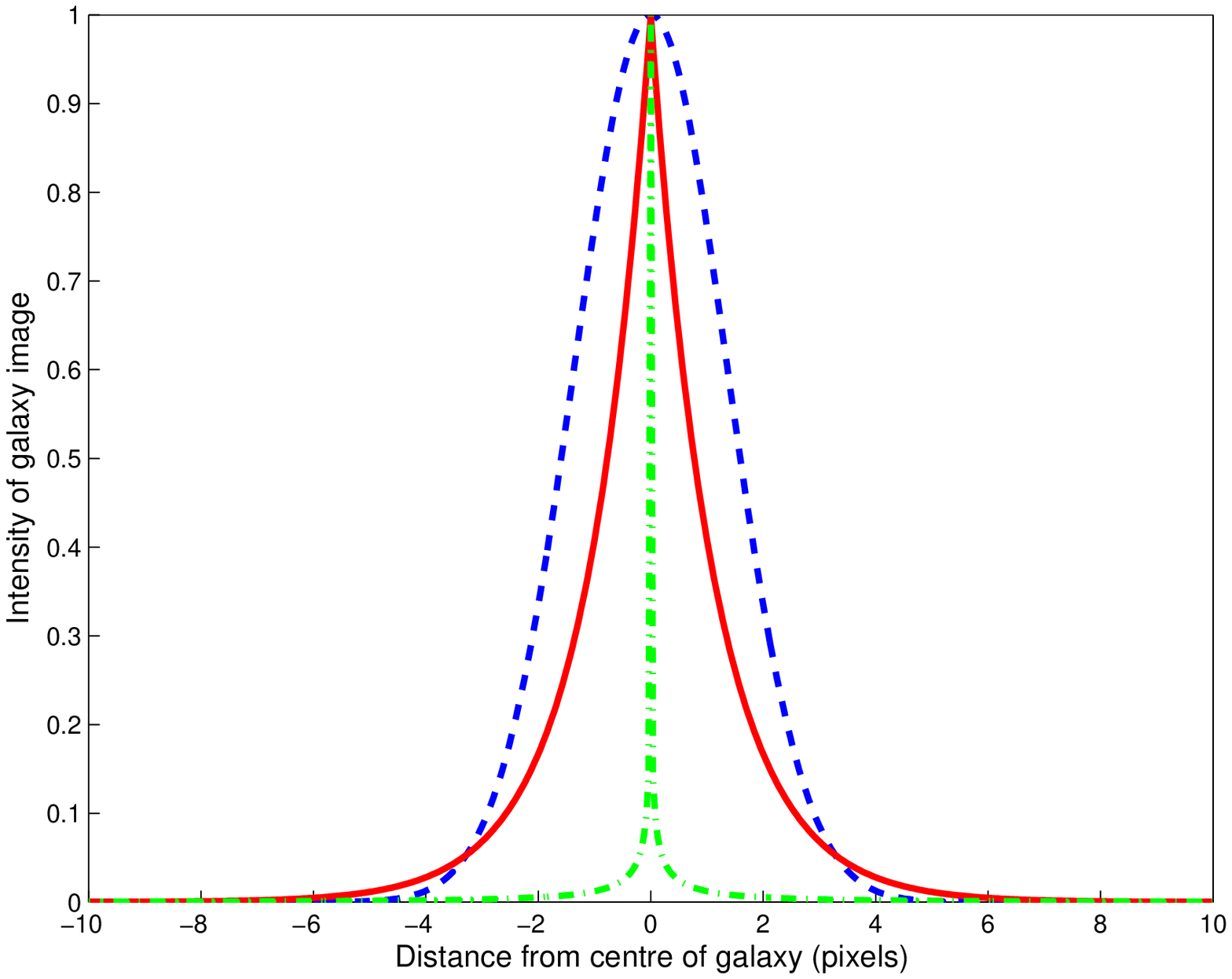,height=6cm,angle=0}
\\
\epsfig{file=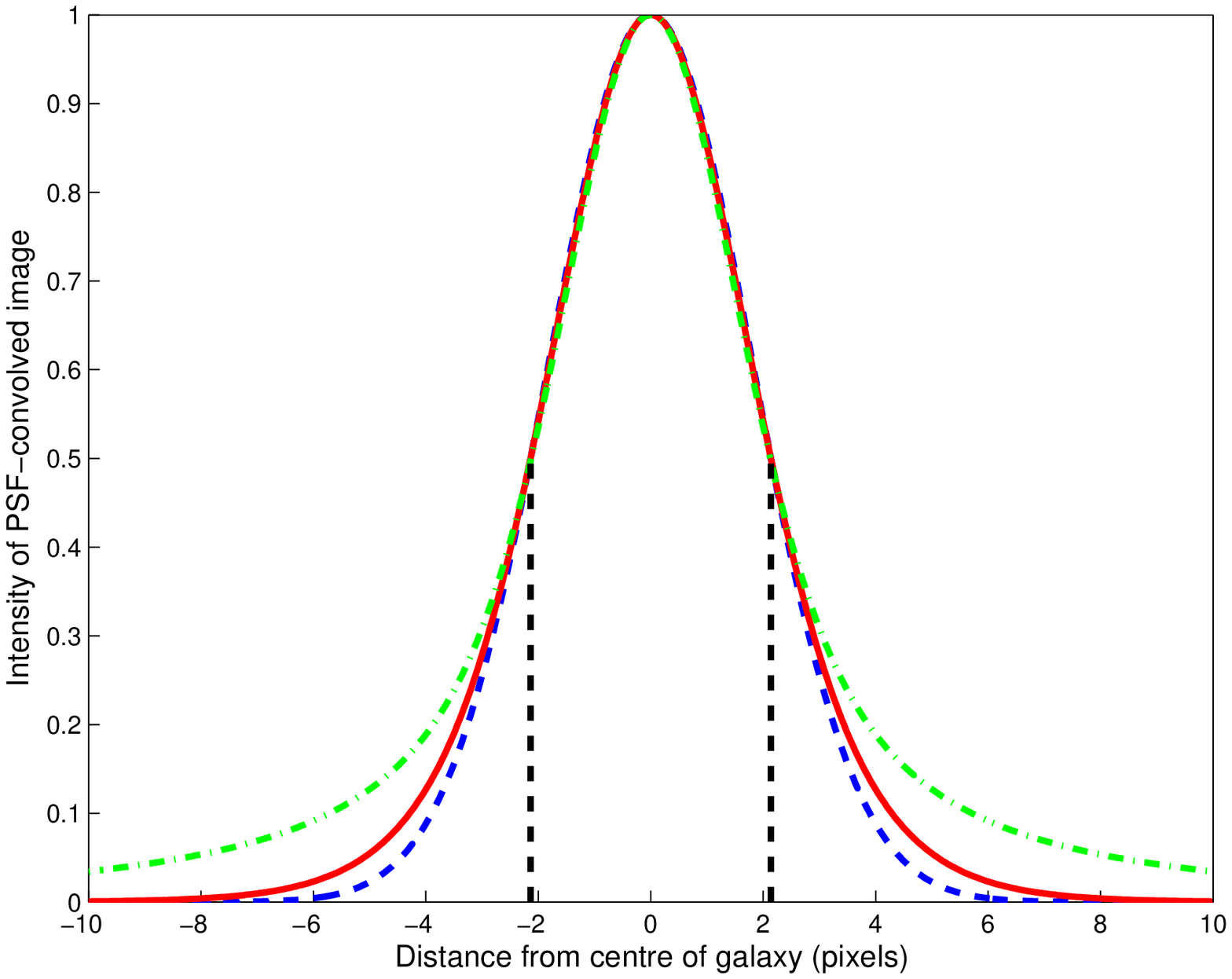,height=6cm,angle=0}
\caption{
Cross-sections through the centre of a Gaussian (blue dashed), exponential (red solid) and de Vaucouleurs (green dash-dot) galaxy (top) and PSF-convolved galaxy (bottom). The PSF is a Gaussian with FWHM equal to 2.85 pixels. The FWHM of the PSF-convolved image (shown by the vertical black dashed lines) is 1.5 times the FWHM of the PSF.
}
\label{fig:reff}
\end{figure}

\section[]{Model fitting using sums of Gaussians}
\label{sect:model}

In this paper we model galaxies as a sum of co-elliptical (homeoidal) Gaussians of varying size and amplitude. This model was first suggested by~\citet{kuijken99}, and developed by~\citet{bridlekbg02} into a publicly available code (im2shape\footnote{http://www.sarahbridle.net/im2shape/}) which has been used to measure cluster masses \citep[e.g.][]{Cypriano04} and tested in the STEP1 simulations~\citep{STEP1}. We stress that the results found in this paper are general for all models adopting elliptical isophotes since any such model can be completely described in terms of a sum of Gaussians. Adopting a sum of Gaussians to model the galaxy has the particular advantage that the convolution with the PSF can be carried out analytically (assuming the PSF is also modelled as a sum of Gaussians).

If the PSF and galaxy intensity profiles are of the form
\begin{eqnarray}
I_{\rm p}(\emph{\textbf{x}})=\frac{
k }{\pi} |\emph{\textbf{C}}_{\rm p}|^{\frac{1}{2}} e^{-k
(\emph{\textbf{x}}-\emph{\textbf{x}}_{0})^T \emph{\textbf{C}}_{\rm p} (\emph{\textbf{x}}-\emph{\textbf{x}}_{0})}
\end{eqnarray}
and
\begin{eqnarray}
I_{\rm g}(\emph{\textbf{x}})=A_{\rm g} e^{-k
(\emph{\textbf{x}}-\emph{\textbf{x}}_{0})^T \emph{\textbf{C}}_{\rm g} (\emph{\textbf{x}}-\emph{\textbf{x}}_{0})}
\end{eqnarray}
respectively, then the PSF-convolved intensity for a sum of $n_{\rm g}$ Gaussians is given by
\begin{eqnarray}
I_{\rm gp}(\emph{\textbf{x}})=\sum_{i=1}^{n_{\rm g}} A_{\rm g, \it i} \frac{|\emph{
\textbf{C}}_{\rm gp, \it i}
|^{\frac{1}{2}}}{|\emph{\textbf{C}}_{\rm g, \it i}|^{\frac{1}{2}}} e^{-k
(\emph{\textbf{x}}-\emph{\textbf{x}}_{0})^T \emph{\textbf{C}}_{\rm gp, \it i} (\emph{\textbf{x}}-\emph{\textbf{x}}_{0})}
\end{eqnarray}
where
\begin{eqnarray}
\emph{\textbf{C}}_{\rm gp, \it i}=\frac{1}{|\emph{\textbf{C}}_{\rm p}+\emph{\textbf{C}}_{\rm g, \it i}|} (|\emph{\textbf{C}}_{\rm p}|\emph{\textbf{C}}_{\rm g, \it i} + |\emph{\textbf{C}}_{\rm g, \it i}|\emph{\textbf{C}}_{\rm p}).
\end{eqnarray}
The centre, ellipticity and orientation of each Gaussian used to model the galaxy are tied. Thus
the number of free parameters in the fit is 4 ($\emph{\textbf{x}}_{0},e,\phi$) plus
$2n_{\rm g}$ ($n_{\rm g}A_{i},n_{\rm g}a_{i}$).
The best-fit parameters are found using $\chi^{2}$-minimisation.
We speed up the calculation by computing the normalisations of the Gaussians analytically. This is possible because the model is linear in these parameters.

Images are generated on a grid $15^{2}$ pixels in size. The intensity in each pixel is the sum of the intensity computed at the centres of $n_{\rm p}^{2}$  sub-pixels, where we refer to $n_{\rm p}$ as the pixel integration level.
%S20090526 [this implies that it should already be obvious?] We note that t
The default pixel integration level used in the simulated galaxies is $n_{\rm p}$=45
(see Section~\ref{sect:sim_gal}).

\section[]{Results for galaxies with elliptical isophotes}
\label{sect:isophotes}

In this Section we simulate galaxies with elliptical isophotes and fit them with different elliptical isophote models. First we try using a single Gaussian when fitting an exponential or de Vaucouleurs profile. We explain our results qualitatively using a one-dimensional toy model. Then we use multiple Gaussians to allow a more accurate fit to the simulations.

\subsection{Using the wrong elliptical isophote model}

\begin{figure}
\center
\epsfig{file=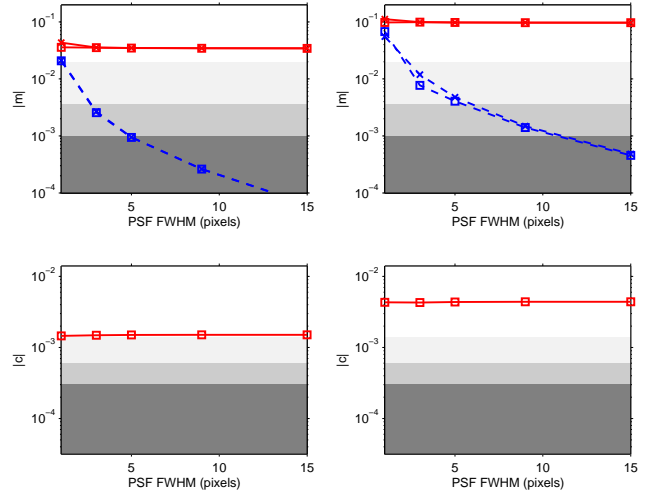,width=8.5cm,angle=0}
\caption{
Multiplicative (top) and additive (bottom) biases for exponential (left) and de Vaucouleurs (right) profiles  as a function of the number of pixels inside the FWHM of the PSF both with (red solid) and without (blue dashed) the PSF included.
Open squares (crosses) show $m_1$,$c_1$ ($m_2$,$c_2$). Any values of $m_i$ and
$c_i$ not seen in the plot lie below the minimum value on the $y$-axis.
The upper edge of each  shaded region (from bottom to top) shows the upper limit on the bias requirements for general far-future, mid-term and upcoming surveys respectively. The survey parameters are given in Table~\ref{tab:requirements}.
The lower limit on the $y$-axis is an order of magnitude less than the requirements for the far-future survey. The top unshaded region is shown for clarity.
}
\label{fig:subpix}
\end{figure}

We first ask whether model fitting using a single Gaussian provides an unbiased shear estimate for a galaxy with elliptical isophotes. We use two different profiles to simulate the true galaxy shape: a de Vaucouleurs and an exponential. The
default model for the PSF is a single Gaussian aligned along the $x$-axis with
perfectly known ellipticity and size. We first investigate how shear measurement biases vary with the size of the pixels used for the observation when the wrong elliptical isophote model is used. We calculate the biases both with the default PSF model and with the PSF set to a delta function. The
default value used in this paper for the PSF FWHM is 2.85 pixels, but
in Fig~\ref{fig:subpix} we vary the resolution from 1 to 15 pixels per PSF FWHM,
%LMV2704
%LMV2704 in we also increase the resolution to 15 pixels per PSF FWHM,
while keeping the relative size of the galaxy and PSF the same.
For the case where the PSF is a delta function the galaxy size is set equal to that computed for the default PSF model (thus the galaxy size is the same at each point on the $x$-axis in Fig~\ref{fig:subpix}).

We ensure that the resolution is the only quantity which changes as the PSF FWHM is increased. This is achieved by convolving the galaxy with the PSF on a large, fine grid and then binning the pixels to obtain images with decreasing resolution. The convolution is carried out as described in Section~\ref{sect:sim_gal}, on a grid $(25\times45)^2$ pixels in size, except here the PSF FWHM is sampled by 45 pixels
instead of ($2.85\times45$) pixels.
The grid is then binned by a factor of 3 (5,9,15,45) to obtain an image in which the PSF FWHM is 15 (9,5,3,1) pixels. The pixel integration level used in each pixel in the galaxy and PSF images prior to the convolution is 1, thus each binned PSF-convolved image has a pixel integration level equal to the binning factor. At each PSF resolution we use the same pixel integration level in the galaxy model as used in the simulated galaxy image.

The dashed lines in Fig~\ref{fig:subpix} show the results for a delta function PSF. The shaded regions show the requirements on $m_i$ and $c_i$ given in Table~\ref{tab:requirements}. The upper edge of each shaded region (from bottom to top) shows the upper limit on the bias allowed for far-future, mid-term and upcoming surveys respectively. The additive shear calibration biases $c_1$ and $c_2$ are always zero when no PSF is used. This is not surprising since there is no preferred direction in which the shear could be biased, since the galaxy direction has been averaged out in the ring-test. The pixels do impose a preferred orientation to the image, but any biases would be the same along the $x$ and $y$ axes and thus positive and negative biases to $c_1$ and $c_2$ are expected to cancel.
We see $m_1$ and $m_2$ decrease as the resolution increases, falling well below forseeable observational requirements (upper edge of grey band).
We discuss the cause of the bias for low resolution images below.
These results indicate that in the limit of infinitely small pixels, model-fitting using a single Gaussian provides an unbiased estimate of the shear of any two-dimensional profile with constant ellipticity isophotes. This agrees with the more general result found by \citet{lewisclt09} that, for the case where there is no PSF, any (wrong) galaxy model will provide unbiased results.

The solid lines in Fig.~\ref{fig:subpix} show the results when the PSF is included. The biases are now significant, independent of the pixel size. In particular, $c_1$ is significant, \emph{even though the PSF model is known precisely}.
This is because in general the angle between the PSF and the galaxy is different for each galaxy in a pair in which the $e_1$ components cancel.
The ring-test
%LMV2105 is
may be %LMV2105
constructed so that $c_2$ is
close to %LMV2105
zero. This provides a useful check on our method. This is simply achieved by aligning the PSF along the $x$-axis and including the mirror image of each galaxy pair in the $y$-axis.
This ensures that the $e_2$ components cancel for image pairs in which the angle between the PSF and the galaxy is the same (except for the galaxy pair at 0 and 90 degrees).

%LMV0705 [inserted paragraph break]
The biases for the de Vaucouleurs profile (right hand panel) are larger than for the exponential profile, which is not surprising considering that it is even further from the single Gaussian used in the fit. %SLB1404
Inserting the bias values into Eq.~\ref{eqn:q} for the exponential galaxy simulation (left-hand panel)
% roughly reading off the plot..
% m_1=0.04; m_2=0.04; c_1=0.002; c_2=0; sigma_gamma=0.03;
% Q = 1e-4 / ( 0.5 * ( (m_1^2 * sigma_gamma^2 + c_1^2) + (m_2^2 * sigma_gamma^2 + c_2^2) ) )
gives $Q\sim 30$, and for the de Vaucouleurs galaxy gives
% m_1=0.1; m_2=0.1; c_1=0.004; c_2=0; sigma_gamma=0.03;
$Q\sim 6$. %LMV0705 [moved from below to remove paragraph break]

%SLB1404 [added the below]
%LMV0705 Inserting the bias values into Eq.~\ref{eqn:q} for the exponential galaxy simulation (left-hand panel)
% roughly reading off the plot..
% m_1=0.04; m_2=0.04; c_1=0.002; c_2=0; sigma_gamma=0.03;
% Q = 1e-4 / ( 0.5 * ( (m_1^2 * sigma_gamma^2 + c_1^2) + (m_2^2 * sigma_gamma^2 + c_2^2) ) )
%LMV0705 gives $Q\sim 30$, and for the de Vaucouleurs galaxy gives
% m_1=0.1; m_2=0.1; c_1=0.004; c_2=0; sigma_gamma=0.03;
%LMV0705 $Q\sim 6$.

\subsection{A qualitative explanation}

\begin{figure}
\center
\epsfig{file=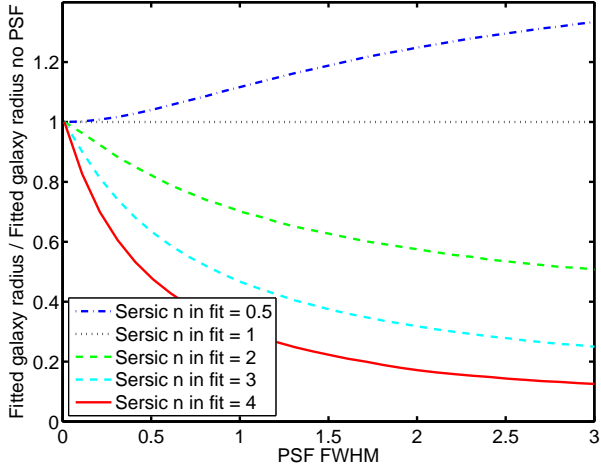,width=8cm,angle=0}
\caption{
Fitted galaxy size as a function of PSF size for a range of fitted
profiles (Sersic index = [ 0.5 1 2 3 4] from top to bottom).
A one-dimensional exponential galaxy was simulated, convolved with a
known Gaussian PSF and fitted with a one-dimensional Sersic profile.
The y axis shows the fitted galaxy size divided by the fitted galaxy
size in the absence of a PSF (PSF FWHM = 0).
}
\label{fig:oned}
\end{figure}

We
%SLB1404 attempt to
explain this result
qualitatively %SLB1404
using Fig~\ref{fig:oned},
which shows results from a toy problem using a one-dimensional image of infinite resolution. %SLB1404
%SLB1404 In this plot
We
%SLB1404 model the galaxy as
simulate a galaxy with %SLB1404
a one-dimensional exponential profile and
%SLB1404 the PSF as a Gaussian.
convolve it with a Gaussian PSF. %SLB1404
The convolved image is then fitted with a Sersic profile convolved with the
correct %SLB1404
PSF.
%SLB1404 (model and parameters known).
The galaxy size (scale radius) is varied to find the best-fit, and this is compared to the best-fit in the absence of a PSF.
The best-fit size of the galaxy is either over-estimated or underestimated, depending on the value of the Sersic index.
%SLB1404, relative to the size found when the PSF is excluded.
The amount by which it is over- or under-estimated increases as the PSF size increases relative to the galaxy.
Fitting a Gaussian galaxy profile (Sersic index = 0.5) causes the fitted galaxy radius to be more overestimated the larger the PSF is, relative to the galaxy size. %SLB1404

%SLB1404 [inserted paragraph break]
Consider now a two-dimensional image of a galaxy with elliptical isophotes aligned along the $x$-axis. Very roughly we can consider biases in the measured ellipticity by considering a one-dimensional slice along the $x$-axis, where the galaxy radius is at its largest relative to the PSF, and then a one-dimensional slice along the $y$-axis where the galaxy radius is at its smallest. %SLB1404
For an elliptical galaxy, therefore,
%SLB1404 we can see from the plot
we expect %SLB1404
that if we use a Gaussian to model the galaxy the size of the major axis will be over-estimated less than the minor axis.
%SLB1404, and the measured ellipticity will be biased low.
This will result in a more circular best fit object, and the shear will be biased low. %SLB1404
By contrast, %SLB1404
if instead we fit the
exponential %SLB1404
galaxy using a de Vaucouleurs profile then, using similar arguments, the
%SLB1404, measured ellipticity, and thus
estimated shear will be biased high.

This conclusion can also be seen qualitatively by considering the two-dimensional image that is being fit. Without the PSF, each point around an elliptical isophote has equal weight in the $\chi^2$, but when the PSF is added, different parts of the galaxy profile are weighted differently.

%SLB1404 [added this paragraph]
In summary, the presence of a convolution causes a bias in the measured shear of an elliptical object, if the wrong profile is assumed.

\begin{figure}
\center
\epsfig{file=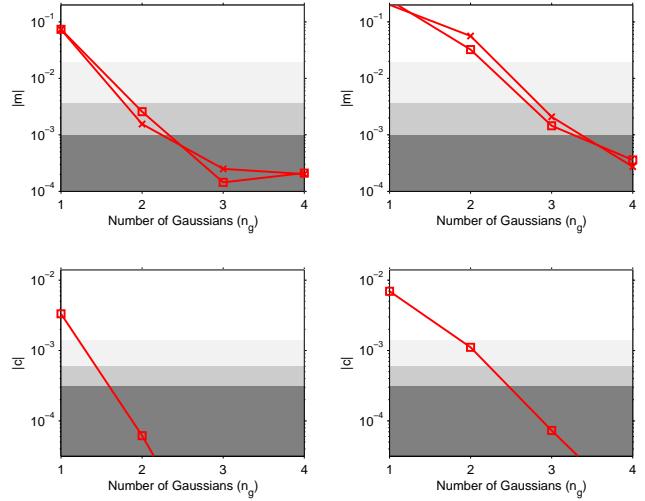,width=8.5cm,angle=0}
\caption{
Multiplicative (top) and additive
(bottom)biases for exponential (left) and de Vaucouleurs (right) profiles
as a function of the number of Gaussians used in the fit. The PSF is included.
The pixel integration level $n_{\rm p}$ is 13. Open squares (crosses) show $m_1$,$c_1$ ($m_2$,$c_2$).
The $c_2$ values are smaller than the minimum on the $y$-axis.
Shaded regions as in Figure~\ref{fig:subpix}.
}
\label{fig:ng}
\end{figure}

\subsection{Allowing the right elliptical isophote model} %SLB1404

We have found that %SLB1404
to obtain an unbiased estimate of the galaxy ellipticity, even when the PSF is known and the pixels are small, the galaxy must be modelled well.
Next %SLB1404
we improve our model by increasing the number of Gaussians used in the sum.
An infinite number of homeoidal Gaussians
%LMV1805 this
would allow perfect reconstruction of any elliptical isophote galaxy. %SLB1404
In Fig~\ref{fig:ng} we show the biases as a function of the number of Gaussians used.
%SLB1404 In this plot the FWHM of the PSF is 2.85 times the pixel size.
We see that the biases reduce
%SLB1404 as the model improves.
to below far-future requirements for both galaxy profiles when 4 Gaussians are used. %SLB1404
For galaxies with an exponential profile only 3 Gaussians are required in the sum.
Note that we do not tune practical computational parameters (especially number of sub-pixels used for pixel integration) for points which already lie well below the requirements for future surveys (darkest shaded area).

\begin{figure}
\center
\epsfig{file=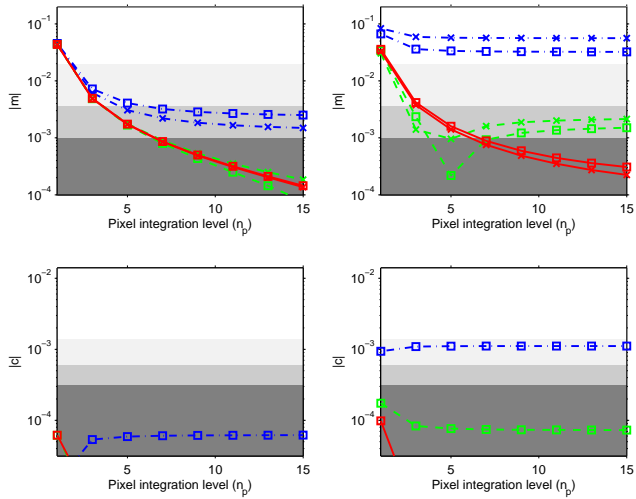,width=8.5cm,angle=0}
\caption{
%LMV2105 Plot showing $\emph{\textbf{m}}$
Multiplicative %LMV2105
(top) and
%LMV2105 $\emph{\textbf{c}}$
additive %LMV2105
(bottom)
biases %LMV2105
for
%LMV2105 an
exponential (left) and de Vaucouleurs (right)
profiles %LMV2105
as a function of the pixel integration level,
%LMV2105 $n_{\rm pix}$.
$n_{\rm p}$. %LMV2105
%LMV2105 , used in the fit (each pixel in the image is the sum of the intensity from $n_{\rm pix}^{2}$ sub-pixels).
%LMV2105 Cyan dotted,
Blue dot-dash, green dashed and red solid curves
%LMV2105 are for
show the biases when
%LMV2105 1,
2, 3 and 4 Gaussians
%LMV2105 respectively
are included in the model %LMV2105
respectively.
Results for 1 Gaussian are larger than the maximum value on the $y$-axis.
The PSF is included. Squares (crosses) show $m_1$,$c_1$ ($m_2$,$c_2$). Shaded regions as in Figure~\ref{fig:subpix}.
}
\label{fig:nifit}
\end{figure}

In Fig~\ref{fig:nifit}
%SLB1404 where
we plot the biases as a function of
the number of sub-pixels used in the %SLB1404
pixel integration.
The $x$-axis shows the number of sub-pixels
$n_{\rm p}$ %LMV2105
in one direction, so the pixel integration sums over values in
%LMV2105 $n_{\rm pix}^2$
$n_{\rm p}^2$ %LMV2105
sub-pixels. %SLB1404
Recall that the default value used e.g. in Fig~\ref{fig:ng} was
%LMV1805 $n_{\rm pix} =15$. %SLB1404 [looks about right!?! check!]
$n_{\rm p} =13$. %LMV1805
Specifically, the biases flatten when limited by the number of Gaussians, and decrease when limited by the pixel integration level.
%SLB1404 We note that i
If  a
%SLB1404 low value is used for the pixel integration
small number of sub-pixels are used in the fit %SLB1404
then the galaxy is more elliptical than in the unpixellated case. This results in an estimated ellipticity which is rounder than the true ellipticity. This effect however cancels out in the ring-test and the decrease in bias with increasing pixel integration is entirely a result of the improvement in the
%SLB1404 galaxy
pixel %SLB1404
model.
We see that
%LMV2105 $n_{\rm pix}\sim10$
$n_{\rm p}\sim10$ %LMV2105
is more than sufficient for forseeable future surveys, and
%LMV2105 $n_{\rm pix}\sim5$
$n_{\rm p}\sim5$ %LMV2105
is sufficient for mid-term surveys. However
%LMV2105 $n_{\rm pix}\sim1$
$n_{\rm p}\sim1$ %LMV2105
is insufficient even for current surveys. %SLB1404

\section[]{Results for bulge plus disk galaxies}
\label{sect:bd}

%SLB1404 [moved 2 paragraphs from here into a new section 3.3]

%SLB1404 [added the following paragraph]
So far all our simulated galaxies have had elliptical isophotes. However this is not the case in the universe, and the simple deviation we consider in this paper is a two-component bulge plus disk model. In Section~\ref{sec:two_compt_models} we described two fiducial two-component models, one to model a spiral galaxy with a bulge, and one to model an elliptical galaxy with a small disk.
We repeat the previous shear measurement bias analysis, always using
an elliptical isophote model in the fit, despite the non-elliptical isophotes of the simulated images. The purpose is to see whether elliptical isophote models can be used for shear measurement from non-elliptical isophote galaxies.

\begin{figure}
\center
\epsfig{file=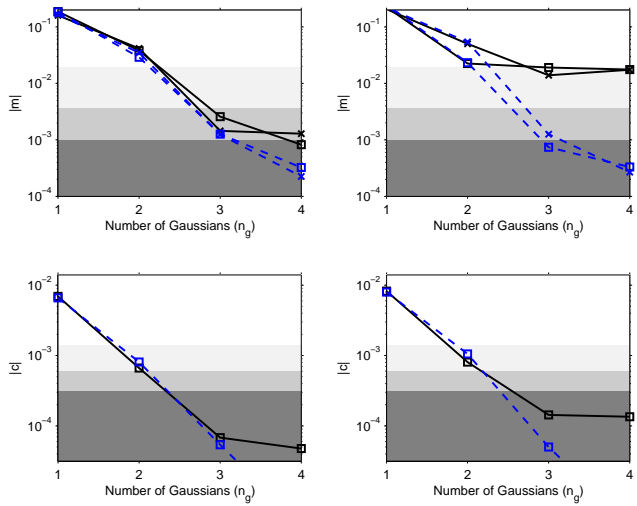,width=8.5cm,angle=0}
\caption{
%LMV2105 Plot showing $\emph{\textbf{m}}$
Multiplicative %LMV2105
(top) and
additive %LMV2105
%LMV2105 $\emph{\textbf{c}}$
(bottom)
biases %LMV2105
for
two-component %LMV2105
%LMV2105 bulge plus disk
galaxies with $r_{\rm d}/r_{\rm b}$ equal to 0.5 (left) and 7.5 (right) as a function of the number of Gaussians used in the fit. The S\'{e}rsic index of the bulge is 4 (left) and 1.5 (right) and in both cases the disk is an exponential. Blue dashed and black solid lines show results for the case where the bulge and the disk have the same ellipticity ($e_{\rm b}=e_{\rm d}=0.2$) and different ellipticities ($e_{\rm b}=0.05$, $e_{\rm d}=0.2$) respectively. The bulge to total flux ratio is 0.8 (left) and 0.3 (right).
Shaded regions as in Figure~\ref{fig:subpix}.
}
\label{fig:bt_ng}
\end{figure}

In Fig.~\ref{fig:bt_ng} we plot the biases for both
two-component %SLB1404
models as a function of the number of (co-elliptical) Gaussians used
%SLB1404 to model the galaxy
in the fit.
%SLB1404 for $B/T=0.3$.
For reference, we also show the results when the bulge ellipticity is equal to the disk ellipticity
%LMV2105 ($e_{\rm bulge}=e_{\rm disk}=0.2$),
($e_{\rm b}=e_{\rm d}=0.2$), %LMV2105
i.e. the simulated galaxy has elliptical isophotes. %SLB1404
When the bulge and disk ellipticity are the same the biases decrease as the number of Gaussians used in the fit increases.
This type of behaviour was already seen in Fig.~\ref{fig:ng}, and the results are slightly different now due to the different galaxy profile arising from the sum of exponential and de Vaucouleurs components.

When the bulge and disk have different ellipticities, however, the bias is not reduced by increasing the number of Gaussians beyond %LMV2105 $n_{\rm G}=3$.
$n_{\rm g}=3$. %LMV2105
We have checked that
%LMV2005 this
this bias is not due to the finite resolution used for %SLB1404
%SLB1404 This bias is not limited by
the pixel integration.
%SLB1404 [added sentence break] but by
We conclude that it is %SLB1404
the failure of the model to take account of galaxies with varying ellipticity isophotes.

We next investigate how the size of the bias depends on the amount of flux in each component. %SLB1404
In Fig~\ref{fig:bt_flux} we plot the biases as a function of the bulge-to-total flux ratio for the spiral and elliptical galaxy models for
%LMV2105 $n_{\rm G}=4$.
$n_{\rm g}=4$. %LMV2105
Again, we include a reference curve for the case where the bulge and disk ellipticity are equal.
As expected, the biases fall to the residual level as $B/T$ approaches zero or unity.
The biases differ from the reference curve for $B/T=1$ because the bulge ellipticity is 0.05 for the solid curve but 0.2 for the dashed (reference) curve. %LMV1805 [added]
The elliptical-like galaxy (left panel) has negligible additive biases, and has multiplicative biases below the requirements of upcoming mid-term surveys at all bulge-to-total ratios.
The behaviour at $B/T=0.7$ is due to a change in sign of $m_i$ from negative at lower $B/T$ values to positive at higher $B/T$ values. %LMV2605

For the spiral galaxy model
%SLB1404 $m$ and $c$
both additive and multiplicative biases %SLB1404
peak at $B/T \sim 0.2$.
The multiplicative bias at this $B/T$ is worse even than the requirements for upcoming surveys. The additive bias is slightly above the requirements for far-future surveys. %SLB1404
Most disk galaxies have $B/T<1/3$~\citep{kormendy08}, with a median value of 0.24 for early-type spiral galaxies (Sa-Sb) and 0.04 for late-type spiral galaxies (Scd-Sm)~\citep{GrahamWorley}.
It is likely that on averaging over all galaxy types the biases are lower than the requirements for upcoming surveys. However,
the exact bias for any particular survey will need to be calculated incorporating the galaxy selection criteria and point spread function.

\section{Discussion}
\label{sect:discussion}

To fully capitalise on the potential of gravitational lensing as a cosmological probe
biases on galaxy shear estimates must be reduced to the sub-percent level. In this paper we have shown that the effects of convolution with the PSF makes this a non-trivial problem. In particular, the unlensed galaxy must be very accurately modelled \emph{even if the PSF is known precisely} and the pixels are small.
%SLB20090518 [added the below sentence]
We have isolated this effect by restricting our investigation to noise-free images.

%SLB20090518 [added below paragraph]
We have illustrated that fitting a single elliptical Gaussian to an elliptical exponential or de Vaucouleurs profile causes no bias on the measured shear, in the
unrealistic case where the pixels are infinitely small and there is no PSF.
For the fiducial galaxy size we chose, application of a realistic PSF causes
a significant shear measurement bias, too large even to use single-Gaussian fitting for current cosmic shear data.
This illustrates the general point that even if galaxies have elliptical isophotes, a model-fitting method must use a realistic galaxy profile. We explained this qualitatively by considering a one-dimensional toy model.

\citet{lewisclt09} proved that the presence of a PSF will result %LMV2105
in biased shear estimates when the wrong galaxy model is used. In this paper we have quantified the level of the bias when the wrong model used
is a sum of co-elliptical %LMV2605
Gaussians, but stress that
our results are general for any model-fitting method using elliptical
profiles.
We find that if galaxies have elliptical isophotes then a sum of 4 Gaussians is sufficient for future surveys. For bulge plus disk galaxies
increasing the number of Gaussians in the model beyond $\sim3$ does not significantly reduce the biases.

Earlier versions of LensFit\footnote{The latest LensFit version fits a co-elliptical bulge plus disk model. See http://www.physics.ox.ac.uk/lensfit/}~\citep{millerkhhv07,kitchingmhvh08} used a de Vaucouleurs profile
to fit galaxies of all types, including exponentials. Thus this is expected to lead to a small residual bias. We found that using an overly flat profile (Gaussian) the shears were biased low relative to the truth. Our toy model predicts that fitting an overly peaky profile (e.g. a de Vaucouleurs to an exponential) will overestimate the shears.

Im2shape~\citep{bridlekbg02} %LMV2105
fits a sum of co-elliptical Gaussians, however there is usually no strong prior on the relative sizes and amplitudes of the components. Therefore when applied to noisy data it is possible that they might not sum to make a particularly peaky profile, and may produce results closer to those expected from fitting a single Gaussian. This could be rectified by applying priors to the relative sizes and amplitudes of the Gaussians, however for best results these priors should be tuned to the expected profiles in the data.

\begin{figure}
\center
\epsfig{file=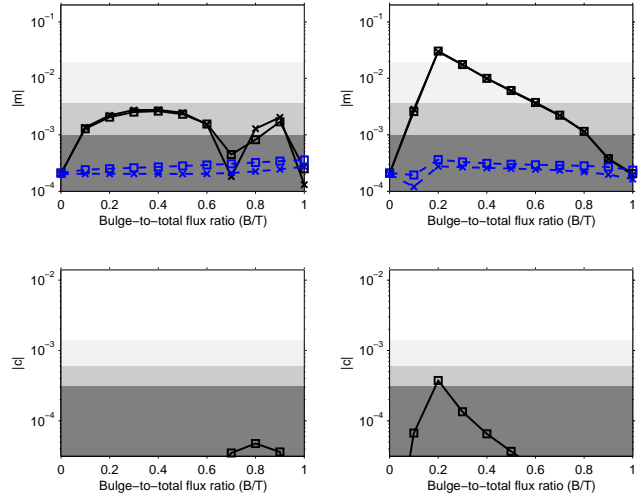,width=8.5cm,angle=0}
\caption{
Multiplicative (top) and additive (bottom) biases for elliptical (left) and spiral (right) two-component galaxies as a function of the bulge-to-total flux ratio.
Black solid and blue dashed curves as in figure~\ref{fig:bt_ng}. Shaded regions as in Figure~\ref{fig:subpix}.
}
\label{fig:bt_flux}
\end{figure}

This result may also be relevant for shapelets methods, which are based on a Gaussian. If only a low order shapelet expansion is used then the profile will be less centrally peaked, and have smaller wings, than an exponential or de Vaucouleurs. A similar expansion based on the sech function has been proposed to address these problems (van Uitert \& Kuijken in prep).

Model-fitting techniques adopting co-elliptical profiles~\citep{bridlekbg02,kuijken06,millerkhhv07,kitchingmhvh08}
cannot, by definition, provide an exact fit to multi-component galaxies with varying ellipticity isophotes.
We find
that this introduces %SLB20090518
a fundamental limit to the accuracy of these methods which
can produce biases on shear measurements from individual galaxies which
are too large for future surveys. The size of the bias depends on the true galaxy morphology, and we investigate just two example morphologies over a range in bulge-to-disk
flux ratios. The bias is largest for spiral-like galaxies with about 20 per cent of the flux in a bulge component. The precise impact on future surveys would require a detailed modelling of galaxy properties and the survey selection function, and is beyond the scope of this work. %SLB20090518
Further, it may be possible to use fudge parameters which correct for the biases resulting from model-fitting with elliptical profiles. It is unclear at this stage how well this would work
given the wide range of underlying galaxy morphologies.

The galaxies simulated in GREAT08 are comparable to the model used in this paper to represent ellipticals.
In addition, the PSF model we use (a single Gaussian) has a similar shape to a Moffat profile with $\beta=3$, used in GREAT08. Further, we adopt the same PSF FWHM (in pixels) and the galaxy to PSF size ratio is close to the central value in the GREAT08 simulations. %LMV2605
From the left hand panel of Fig.~\ref{fig:bt_flux} we find
$m\sim3
\times10^{-3}$, $c\sim3\times10^{-5}$
which would give a GREAT08 $Q$
of $\sim7000$. This is indicative of an upper limit to the
GREAT08 $Q$ obtainable by shape measurement techniques using model-fitting with elliptical isophotes.

We note that although the simple %SLB20090518
bulge plus disk galaxies
considered here %SLB20090518
are only an approximation to real systems which often contain more than two structural components, such as nuclear sources, bars, spiral arms, and H\texttt{II} regions, the results we obtain
provide an illustration of the level of bias that may be incurred, and show that more detailed simulations would be required to test elliptical isophote model-fitting methods for future surveys.
In addition, further investigation is required to quantify the bias level for various (survey-dependent) PSF models (e.g. models including extended wings, dipole moments etc). %LMV2605
Simulations incorporating complex galaxy and PSF models are %LMV2605
%LMV2605 This is
anticipated for some GREAT Challenges in the future.

Stacking many galaxies in a similar region of sky should circumvent the dependence on individual galaxy properties, as suggested by ~\citet{kuijken99} and~\citet{lewisclt09}. If we are interested only in some average shear for these galaxies then this may be measured from the stacked image, from which detailed galaxy substructure will have been washed out, to leave an elliptical object with an ellipticity corresponding to the average shear (in the limit of an infinite number of averaged galaxies). This approach now requires more detailed study to determine its practical feasibility.

We have shown that the underlying galaxy shape must be accurately modelled to obtain unbiased shear estimates.
However, considerable information about the galaxy shape is lost when images are pixellated and noise added.
The optimal freedom in the model may be determined by a balance which allows the model to account for the wide range of galaxy morphologies while restricting it from fitting to noise spikes.
Future shape measurement methods should capitalise on the wealth of knowledge gathered in the field of galaxy shape classification. Information about, for example, the narrow range in bulge-to-disk size ratios observed in spirals~\citep{GrahamWorley} could be fed into shape measurement methods using a Bayesian approach. Such methods would need to be fine-tuned for different surveys.

\section*{Acknowledgments}
We thank Antony Lewis, John Bridle, Tom Kitching, Jean-Paul Kneib, Alexandre Refregier, Adam Amara, Stephane Paulin-Henrikkson, Phil Marshall, Konrad Kuijken, Gary Bernstein, Eduardo Cypriano, Benjamin Joachimi and Steve Gull for useful discussions.
SLB thanks the Royal Society for support in the form of a University
Research Fellowship. LMV acknowledges support from the STFC.

\bsp

\bibliographystyle{mn2e}
%\bibliography{bibfile}

\label{lastpage}

\end{document}